\documentclass{webofc}

\usepackage[varg]{txfonts}   
\begin{document}
\title{Exclusive semileptonic $\mathbf{B}$-meson decays using lattice QCD and unitarity}
%
%

\author{\firstname{Guido} \lastname{Martinelli}\inst{1}\fnsep\thanks{\email{guido.martinelli@roma1.infn.it}} \and
        \firstname{Manuel} \lastname{Naviglio}\inst{2,3}\fnsep\thanks{\email{manuel.naviglio@phd.unipi.it}} \and
        \firstname{Silvano} \lastname{Simula}\inst{4}\fnsep\thanks{\email{silvano.simula@roma3.infn.it}} \and
        \firstname{Ludovico} \lastname{Vittorio}\inst{3,5}\fnsep\thanks{\email{ludovico.vittorio@sns.it}}
}

\institute{Physics Department, University of Rome "La Sapienza" and INFN, Sezione di Roma, Piazzale Aldo Moro 5, I-00185 Rome, Italy 
\and
           Physics Department, University of Pisa, Largo Bruno Pontecorvo 3, I-56127 Pisa, Italy 
\and
           INFN, Sezione di Pisa, Largo Bruno Pontecorvo 3, I-56127 Pisa, Italy
           \and
           INFN, Sezione di Roma Tre, Via della Vasca Navale 84, I-00146 Rome, Italy
           \and
           Scuola Normale Superiore, Piazza dei Cavalieri 7, I-56126 Pisa, Italy
          }

\abstract{%
  We present the results of the application of the Dispersion Matrix approach to exclusive semileptonic $B$-meson decays. This method allows to determine the hadronic form factors in a non-perturbative and completely model-independent way. Starting from lattice results available at large values of the momentum transfer, the behaviour of the form factors in their whole kinematical range is obtained without introducing any parameterization of their momentum dependence. We will focus on the determination of the Cabibbo-Kobayashi-Maskawa matrix elements $\vert V_{cb} \vert$ and $\vert V_{ub} \vert$ through the analysis of $B_{(s)} \to D_{(s)}^{(*)} \ell \nu$ and $B_{(s)} \to \pi(K) \ell \nu$ decays. New theoretical determinations of the Lepton Flavour Universality ratios relevant for these transitions will be also presented. }
\maketitle
\section{State-of-the-art of exclusive semileptonic $\mathbf{B}$-meson decays}
\label{intro}
Exclusive semileptonic $B$-meson decays are one of the most challenging processes in the phenomenology of flavour physics, since they are affected by two unsolved problems. 

On the one hand, we have the so-called $\vert V_{cb} \vert$ \emph{puzzle}, $i.e.$ the discrepancy between the inclusive and the exclusive determinations of the CKM matrix element $\vert V_{cb}\vert$. According to the FLAG Review 2021 \cite{FLAG21}, there is a $\sim2.8\sigma$ tension between the exclusive estimate (which depends on the form factors parametrization) and the inclusive one, namely
\begin{equation}
\label{VcbFLAG21}
\vert V_{cb} \vert_{\rm excl} \times 10^3 = 39.36(68),\,\,\,\,\,\,\,\,\,\,\,\,\vert V_{cb} \vert_{\rm incl} \times 10^3 = 42.00(65).
\end{equation}
Two new estimates of the inclusive value have also recently appeared, namely $\vert V_{cb} \vert_{\rm incl} \times 10^3=42.16(50)$ \cite{Bordone:2021oof} and $\vert V_{cb} \vert_{\rm incl} \times 10^3=41.69(63)$ \cite{Bernlochner:2022ucr}, which are compatible with the inclusive FLAG value in Eq.\,(\ref{VcbFLAG21}) and corroborate its truthfulness. 

On the other hand, a strong tension exists between the theoretical values and the measurements of $R(D^{(*)})$, which are a fundamental test of Lepton Flavour Universality (LFU) and are defined as 
\begin{equation}
\label{RDdef}
R(D^{(*)}) \equiv \frac{\Gamma(B \to D^{(*)} \tau \nu_{\tau})}{\Gamma(B \to D^{(*)} \ell \nu_{\ell})},
\end{equation}
where $\ell = e, \mu$ denotes a light lepton. The HFLAV Collaboration \cite{HFLAV} has recently computed the world averages of the available measurements of the $R(D^{(*)})$ ratios and of their SM theoretical expectations, obtaining 
\begin{equation}
\label{RDHFLAV}
R(D)_{\rm SM} = 0.299 \pm 0.003, \qquad R(D)_{\rm exp} = 0.339 \pm 0.026 \pm 0.014
\end{equation}
for the $B \to D$ case and 
\begin{equation}
\label{RDstHFLAV}
R(D^*)_{\rm SM} = 0.254 \pm 0.005, \qquad R(D^*)_{\rm exp} = 0.295 \pm 0.010 \pm 0.010
\end{equation}
for the $B \to D^*$ one. As clearly stated by HFLAV Collaboration, the averages of the measurements of $R(D)$ and $R(D^*)$ exceed the corresponding SM predictions by 1.4$\sigma$ and 2.8$\sigma$, respectively. If the experimental correlation between these two quantities, namely $\rho=-0.38$, is taken into account, the resulting difference with the SM is increased to the 3.3$\sigma$ level \cite{HFLAV}.

\section{The Dispersion Matrix approach to hadronic Form Factors}
\label{Section2}
Let us now focus on semileptonic $B \to D^{(*)}$ decays for massless leptons (namely $\ell = e, \mu$ in what follows). In case of production of a \emph{pseudoscalar} (PS) meson, $i.e.$ the $B \to D \ell \nu$ case, the differential decay width reads
\begin{equation}
\label{finaldiff333}
\frac{d\Gamma}{dq^2}=\frac{G_F^2}{24\pi^3} \eta_{EW}^2 \vert V_{cb} \vert^2 \vert \vec{p}_{D}\vert^3  \vert f^+(q^2) \vert^2,
\end{equation}
where $\vec{p}_{D}$ represents the 3-momentum of the produced $D$ meson. In this case, there is only one Form Factor (FF) appearing in the theoretical expression of the decay width, namely $f^+(q^2)$. In case of production of a \emph{vector} meson, $i.e.$ the $B \to D^* \ell \nu$ case, the expression of the differential decay width is instead
\begin{equation}
\begin{aligned}
\label{finaldiff333BDst}
\frac{d\Gamma(B \rightarrow D^{*}(\rightarrow D\pi) \ell \nu)}{dw d\cos \theta_{\ell} d\cos \theta_v d\chi}& =\frac{3}{4}\frac{G_F^2 }{(4\pi)^4} \eta_{EW}^2 \vert V_{cb} \vert^ 2 m_B^3 r^2 \sqrt{w^2-1} (1+r^2-2rw)  \\
& \times \{ (1-\cos \theta_{\ell} )^2 \sin^2 \theta_v \vert H_{+} \vert^2 + (1+\cos \theta_{\ell} )^2\sin^2 \theta_v \vert H_{-} \vert^2 \\
& + 4 \sin^2 \theta_{\ell}\cos^2 \theta_v\vert H_{0} \vert^2 - 2 \sin^2 \theta_{\ell}\sin^2 \theta_v \cos 2\chi  H_{+}  H_{-} \\
& - 4 \sin \theta_{\ell} (1-\cos \theta_{\ell} ) \sin\theta_v\cos\theta_v\cos\chi H_{+}  H_{0}\\
& + 4 \sin \theta_{\ell} (1+\cos \theta_{\ell} ) \sin\theta_v\cos\theta_v\cos\chi H_{-}  H_{0}\},
\end{aligned}
\end{equation}
where we have defined $r \equiv m_{D^*}/m_B$ and we have introduced the helicity amplitudes 
\begin{equation}
\label{helampl}
H_0(w) = \frac{\mathcal{F}_1(w)}{\sqrt{m_B^2+m_D^2-2m_Bm_Dw}}, \qquad H_{\pm}(w) = f(w) \mp m_B m_{D^*} \sqrt{w^2-1}\,g(w).
\end{equation}
In Eq.(\ref{finaldiff333BDst}) $\theta_l, \theta_v, \chi$ represent the helicity angles and the FFs are $f(w),\,g(w),\,\mathcal{F}_1(w)$. Note that in Eqs.\,(\ref{finaldiff333}) and (\ref{finaldiff333BDst}) we refer equivalently to the momentum transfer $q^2$ or to the recoil $w$, since they are related by 
$q^2(w)= m_B^2+m_{D^{(*)}}^2 - 2 m_B m_{D^{(*)}} w = m_B^2 \, (1+r^2-2rw)$.

Let us finally stress again that in the previous expressions we are assuming a massless produced lepton. For massive leptons ($\ell = \tau$), instead, one should add the FFs $f_0(q^2)$ for semileptonic $B \to D$ decays and $P_1(w)$ for semileptonic $B \to D^*$ one.

Now, our goal is to describe the FFs entering in exclusive semileptonic $B$-meson decays by using the novel Dispersion Matrix (DM) method \cite{DiCarlo:2021dzg}, which was originally proposed in \cite{Lellouch96}. The DM method allows us to study the FFs in a non-perturbative and model-independent way. Starting from the available LQCD computations of the FFs at high momentum transfer (or, equivalently, at low recoil), we can extrapolate their behaviour in the opposite kinematical region without assuming any functional dependence of the FFs on $q^2$ (or, equivalently, on $w$) and using only non-perturbative inputs. Moreover, the resulting bands of the FFs will be independent of the experimental determinations of the differential decay widths. 

From the mathematical point of view, the starting point is to focus on one FF, for instance $f$, and then consider the matrix
\begin{equation}
\label{eq:Delta2}
\mathbf{M} = \left( 
\begin{tabular}{ccccc}
   $\chi$ & $\phi f$ & $\phi_1 f_1$ & $...$ & $\phi_N f_N$ \\[2mm] 
   $\phi f$ & $\frac{1}{1 - z^2}$ & $\frac{1}{1 - z z_1}$ & $...$ & $\frac{1}{1 - z z_N}$ \\[2mm]
   $\phi_1 f_1$ & $\frac{1}{1 - z_1 z}$  & $\frac{1}{1 - z_1^2}$ & $...$ & $\frac{1}{1 - z_1 z_N}$ \\[2mm]
   $... $  & $...$ & $...$ & $...$ & $...$ \\[2mm]
   $\phi_N f_N$ & $\frac{1}{1 - z_N z}$ & $\frac{1}{1 - z_N z_1}$ & $...$ & $\frac{1}{1 - z_N^2}$
\end{tabular}
\right),
\end{equation}
where we have introduced the conformal variable $z$ defined as 
\begin{equation}
\label{conf}
z(t) = \frac{\sqrt{t_+ -t} - \sqrt{t_+ - t_-}}{\sqrt{t_+ -t} + \sqrt{t_+ - t_-}},\,\,\,\,\,\,\,\,\,\,\,\,\,\,\,\,\,\,\,\,\,t_{\pm}=(m_B \pm m_{D^{(*)}})^2
\end{equation}
with $t \equiv q^2$, or, equivalently, as
\begin{equation}
z(w)=\frac{\sqrt{w+1}-\sqrt{2}}{\sqrt{w+1}+\sqrt{2}}.
\end{equation}
In this expression, $\phi_i f_i \equiv \phi(z_i) f(z_i)$ (with $i = 1, 2, ... N$) are the known values of the quantity $\phi(z) f(z)$ corresponding to the values $z_i$ at which the FFs have been computed on the lattice. The kinematical function $\phi(z)$ has a specific expression for each FF \cite{paperoIII}. Finally, the susceptibility $\chi(q_0^2)$ is related to the derivative with respect to $q_0^2$ of  the Fourier transform of suitable Green functions of bilinear quark operators and follows from the dispersion relation associated to a particular spin-parity quantum channel. Note that they have been computed for the first time on the lattice in \cite{paperoII} for $b \to c$ quark transitions by choosing $q_0^2=0$. At this point, one can demonstrate from first principles that $\det \mathbf{M} \geq 0$. Then, the positivity of the determinant, which we will refer to as \emph{unitarity filter} hereafter, allows to compute the lower and the upper bounds of the FF of interest for each generic value of $z$, $i.e.$
\begin{equation}
f_{\rm lo}(z) \leq f(z) \leq f_{\rm up}(z).
\end{equation}
To be more quantitative, we have that \cite{DiCarlo:2021dzg}
\begin{equation}
   \beta - \sqrt{\gamma} \leq  f \leq \beta + \sqrt{\gamma} ~ , ~
    \label{eq:bounds}
\end{equation} 
where (after some algebraic manipulations)
\begin{equation*}
      \beta = \frac{1}{\phi(z)d(z)} \sum_{j = 1}^N f_j \phi_j d_j \frac{1 - z_j^2}{z_0 - z_j},\,\,\,\,\,\,\,\,\,\, \gamma  =   \frac{1}{1 - z_0^2} \frac{1}{\phi^2(z) d^2(z)} \left( \chi - \overline{\chi} \right),   
\end{equation*} 
\begin{equation*}
       \overline{\chi}  =  \sum_{i, j = 1}^N f_i f_j \phi_i d_i \phi_j d_j \frac{(1 - z_i^2) (1 - z_j^2)}{1 - z_i z_j}
\end{equation*}
where $d(z),\,d_i$ are simply kinematical functions \cite{DiCarlo:2021dzg}. Unitarity is satisfied only when $\gamma \geq 0$, which implies $\chi \geq \overline{\chi}$. One can show that the values of $\beta$ and $\gamma$ depend on $z$, while the value of $\overline{\chi}$ does not depend on $z$ and it depends only on the set of input data. Consequently, the unitarity condition $\chi \geq \overline{\chi}$ does not depend on $z$.

In what follows, we will describe the results of the phenomenological applications of the DM method to several exclusive semileptonic $B$-meson decays. We will analyze in detail two instructive cases, $i.e.$ $B \to D^{*} \ell \nu$ and $B \to \pi \ell \nu$ transitions, and then we will give an overview of all the results relevant for phenomenology obtained so far.

\section{The DM application to semileptonic $B \to D^{*} \ell \nu$ decays}

\begin{figure}[htb!]
\centering
\includegraphics[width=11cm,clip]{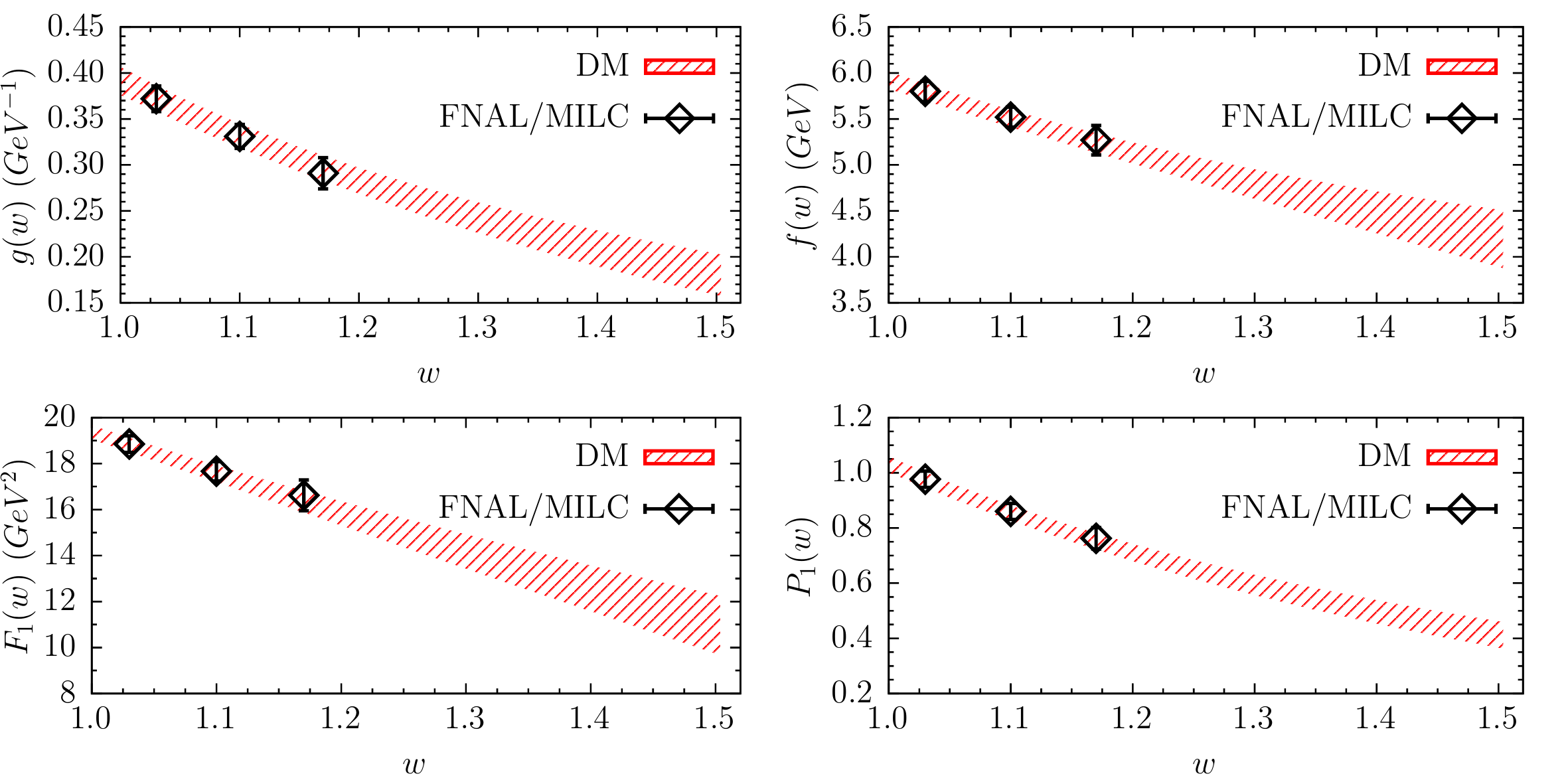}
\vspace{-0.25cm}
\caption{The bands of the FFs $g(w)$, $f(w)$, $\mathcal{F}_1(w)$ and $P_1(w)$ computed by the DM method after imposing both the unitarity filter and the KCs. The FNAL/MILC values~\cite{FNALMILCD*} used as inputs for the DM method are represented by the black diamonds.}
\label{FFD*bis}       
\end{figure}

Let us firstly discuss in detail semileptonic $B \to D^{*}$ decay, which is very challenging due to the high number of FFs involved. In \cite{EPJC} we have computed the unitarity bands of the FFs, starting from the final results of the computations on the lattice performed by the FNAL/MILC Collaborations \cite{FNALMILCD*}. There, in the ancillary files, the authors give the synthetic values of the FFs $g(w), f(w), \mathcal{F}_1(w)$ and $\mathcal{F}_2(w)$ at three non-zero values of the recoil variable $w$, namely $w = \{1.03,1.10,1.17\}$, together with their correlations. Note that the FF $\mathcal{F}_2(w)$ is directly related to the $P_1(w)$ one, in fact $P_1(w) =  \mathcal{F}_2(w) \sqrt{r} / (1 + r)$, where $r \equiv m_{D^*}/m_B \simeq 0.38$. In Fig.\,\ref{FFD*bis} we show the results of our DM analysis as red bands. The DM unitarity bands are built up through bootstrap events that satisfy exactly both the unitarity filter discussed in the previous Section and the Kinematical Constraints (KCs) 
\begin{equation}
    \mathcal{F}_1(1)  =  m_B (1 - r) f(1), \qquad P_1(w_{max})  =  \frac{\mathcal{F}_1(w_{max})}{m_B^2 (1 + w_{max}) (1 -r) \sqrt{r}},
\end{equation}
where $w_{\rm max}$ is the maximum value of the recoil, namely $w_{\rm max} \simeq 1.5$. Starting from the unitary bands shown in Fig.\,\ref{FFD*bis}, we are then able to compute new estimates of both the CKM matrix element $\vert V_{cb} \vert$ and the ratio $R(D^*)$. 

\begin{figure}[htb!]
\centering
\includegraphics[width=11cm,clip]{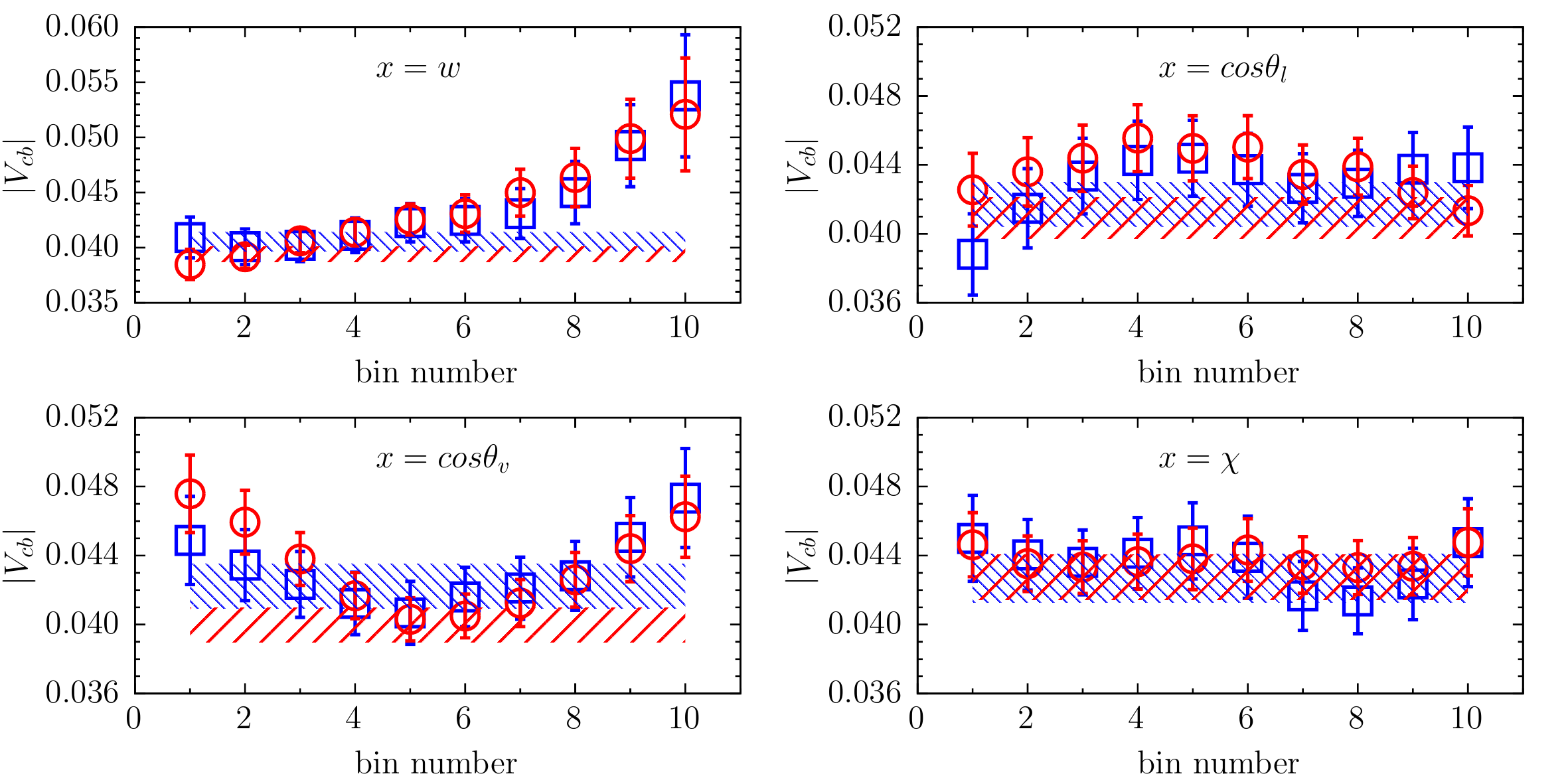}
\vspace{-0.25cm}
\caption{The bin-per-bin estimates of $\vert V_{cb} \vert$ and their weighted means for each kinematical variable ($x=w,\cos\theta_l,\cos\theta_v,\chi$). The blue squares and the red circles correspond respectively to the first \cite{Belluno} and to the second \cite{Venezia} set of the Belle measurements. The dashed blue (red) bands are the results of the average (\ref{muVcbfinal}) in the case of the blue squares (red circles) for each variable .
}
\label{Vcbbinbin}       
\end{figure}

For what concerns $\vert V_{cb} \vert$, the Belle Collaboration \cite{Belluno, Venezia} has published two different sets of measurements of the differential decay widths $d\Gamma/dx$ ($x=w,\cos\theta_l,\cos\theta_v,\chi$) in the form of ten bins for each kinematical variable. Thus, we have computed the theoretical estimate of each one-dimensional differential decay width (starting from the complete expression in the Eq.\,(\ref{finaldiff333BDst}) and using the unitarity bands of the FFs) and we have compared them to the experimental data in order to obtain bin-per-bin estimates of $\vert V_{cb} \vert$.  They are shown in Fig.\,\ref{Vcbbinbin}, where the blue squares come from the first set of Belle measurements \cite{Belluno} while the red circles from the second one \cite{Venezia}. For each kinematical variable and for each experiment, we compute the average of these bin-per-bin estimates as
\begin{equation}
\label{muVcbfinal}
\vert V_{cb} \vert = \frac{\sum_{i,j=1}^{10} (\mathbf{C}^{-1})_{ij} \vert V_{cb} \vert_j}{\sum_{i,j=1}^{10} (\mathbf{C}^{-1})_{ij}}, \qquad \sigma^2_{\vert V_{cb} \vert} = \frac{1}{\sum_{i,j=1}^{10} (\mathbf{C}^{-1})_{ij}},
\end{equation}
where $\mathbf{C}$ is the covariance matrix and $\vert V_{cb} \vert_i$ is the $\vert V_{cb} \vert$ estimate for the $i$-th bin. This procedure generates the dashed blue (red) bands in Fig.\,\ref{Vcbbinbin} for the first (second) set of Belle measurements. We finally combine the resulting eight mean values as
\begin{equation}
\label{sigma28}
\mu_{x} = \frac{1}{N} \sum_{k=1}^N x_k, \qquad \sigma^2_{x} = \frac{1}{N} \sum_{k=1}^N \sigma_k^2 + \frac{1}{N} \sum_{k=1}^N(x_k-\mu_{x})^2,
\end{equation}
obtaining
\begin{equation*}
\vert V_{cb} \vert \times 10^{3} = 41.3 \pm 1.7.
\end{equation*}

The DM method allows us also to obtain fully-theoretical expectation values of other quantities relevant for phenomenology, $i.e.$ the anomaly $R(D^*)$,
the $\tau$-polarization $P_{\tau}$ and finally the $D^*$ longitudinal polarization $F_L$, namely
\begin{eqnarray*}
R(D^*) = 0.275 \pm 0.008,\,\,\,\,\,\,\,\,\,\,P_{\tau} = -0.529 \pm 0.007,\,\,\,\,\,\,\,\,\,\,F_L = 0.414 \pm 0.012.
\end{eqnarray*}

\section{The DM application to semileptonic $B \to \pi \ell \nu$ decays}

The DM method can be applied to any semileptonic charged-current decays of hadrons. In this Section we investigate its potential in the analysis of the $b \to u$ quark transitions by discussing the case of the $B \to \pi \ell \nu$ decays \cite{paperoV}. In this case, since the pion is a PS meson, the formalism is analogous to the one characterizing $B \to D \ell \nu$ decays described in Sec.~\ref{Section2}. 

Thus, $B \to \pi \ell \nu$ transitions are characterized by two FFs:  $f_+^{\pi}(q^2)$ and $f_0^{\pi}(q^2)$. These FFs have been studied by the RBC/UKQCD\,\cite{Flynn:2015mha} and the FNAL/MILC\,\cite{Lattice:2015tia} Collaborations. For both channels the lattice computations of the FFs are available in the large-$q^2$ region\footnote{To be more specific, the authors of Ref.\,\cite{Flynn:2015mha} provide synthetic LQCD values of the FFs (together with their statistical and systematic correlations) at $q^2 = \{ 19.0, 22.6, 25.1 \}$ GeV$^2$. In \cite{Lattice:2015tia}, instead, only the results of BCL fits \cite{BCL} of the FFs extrapolated to the continuum limit and to the physical pion point are available. Thus, from the marginalized BCL coefficients we evaluate the mean values, uncertainties and correlations of the FFs at the same three values of $q^2$ given in Ref.\,\cite{Flynn:2015mha}.}. 

\begin{figure}[htb!]
\centering
\includegraphics[width=0.9\textwidth]{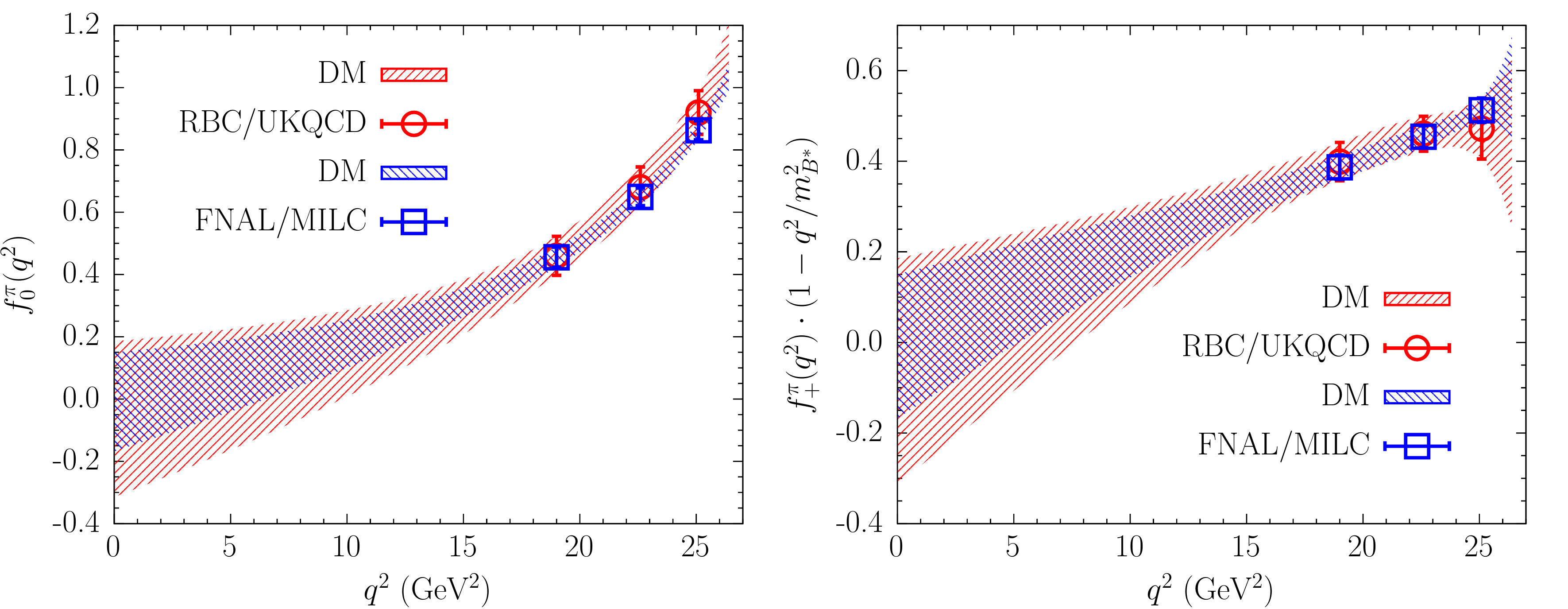}
\vspace{-0.25cm}
\caption{\small The scalar $f_0^\pi(q^2)$ (left panel) and vector $f_+^\pi(q^2)$ (right panel) FFs entering the semileptonic $B \to \pi \ell \nu_\ell$ decays computed by the DM method as a function of the 4-momentum transfer $q^2$ using the LQCD inputs from RBC/UKQCD (red points) and FNAL/MILC (blue squares) Collaborations. In the right panel, the vector FF is multiplied by the factor $(1 - q^2 / m_{B^*}^2)$ with $m_{B^*} = 5.325$ GeV.\hspace*{\fill}}
\label{FFsBpi}
\end{figure}

In Fig.\,\ref{FFsBpi} we show the red (blue) DM bands obtained by using as inputs the RBC/UKQCD (FNAL/MILC) data. Note that using the BCL fits of \cite{Lattice:2015tia} the mean value and the uncertainty of the FFs extrapolated at zero momentum transfer are not stable against variation of the truncation order of the series expansion of the FFs. On the contrary, the DM approach is completely independent of this issue, since no approximation due to the truncation of a series expansion is present. The DM method is equivalent to the results of all possible fits which satisfy unitarity and, at the same time, reproduce exactly the input data. This property is particularly useful in $B \to \pi \ell \nu$ decays, where the extrapolation to $q^2 = 0$ is quite long.

\begin{figure}[htb!]
\centering
\includegraphics[width=0.9\textwidth]{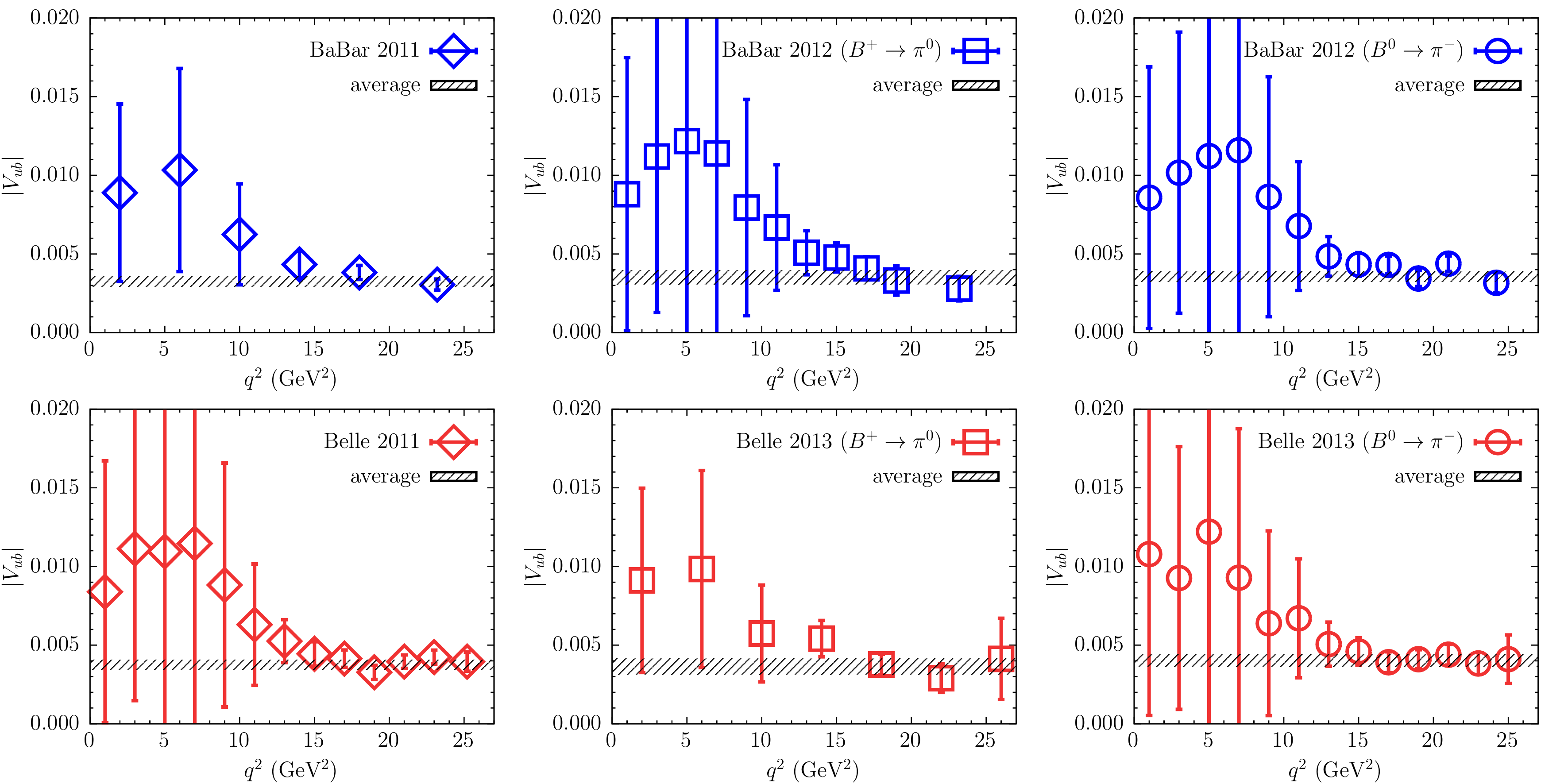}
\vspace{-0.25cm}
\caption{\it \small Bin-per-bin estimates of $\vert V_{ub} \vert$ for each of the six experiments of Refs.\,\cite{delAmoSanchez:2010af, Ha:2010rf, Lees:2012vv, Sibidanov:2013rkk} specified in the insets of the panels as a function of $q^2$. The theoretical DM bands of the FFs correspond to the use of the combination of the available LQCD data as inputs \cite{paperoV}. The (black) dashed bands represent the correlated weighted averages for each experiment.}
\label{VubBpifig}
\end{figure}

For the extraction of the CKM matrix element we compute bin-per-bin values of $\vert V_{ub} \vert$ for each $q^2$-bin of each available experiment. Recall that the branching fractions of this transition have been measured by several experiments \cite{delAmoSanchez:2010af, Ha:2010rf, Lees:2012vv, Sibidanov:2013rkk}. Thus, we evaluate the CKM matrix element for the $n$-th experiment ($n = 1, \ldots, 6$ for the semileptonic $B \to \pi$ decays) through expressions analogous to Eq.\,(\ref{muVcbfinal}). Our results are shown in Fig.\,\ref{VubBpifig} when one uses the combination of the RBC/UKQCD and FNAL/MILC data as inputs of the DM method.

Our final result for $\vert V_{ub} \vert$ is evaluated making use of the averaging procedure given by Eq.\,(\ref{sigma28}) and reads
\begin{equation}
\label{VubLASTCOMB}
\vert V_{ub} \vert^{B\pi} \cdot 10^{3}  =  3.62 \pm 0.47.
\end{equation}
Let us mention here that we are currently investigating strategies to improve the precision of the determination of $\vert V_{ub} \vert$ within our DM approach. Some results can be found in \cite{CKM21}, where our improved determination of the CKM matrix element reads 
\begin{equation}
\label{Vubimpr}
\vert V_{ub} \vert^{B\pi}_{\rm impr} \cdot 10^{3}  =  3.88 \pm 0.32.
\end{equation}

\section{Conclusions}

\begin{figure}[htb!]
\centering
\includegraphics[width=1.\textwidth]{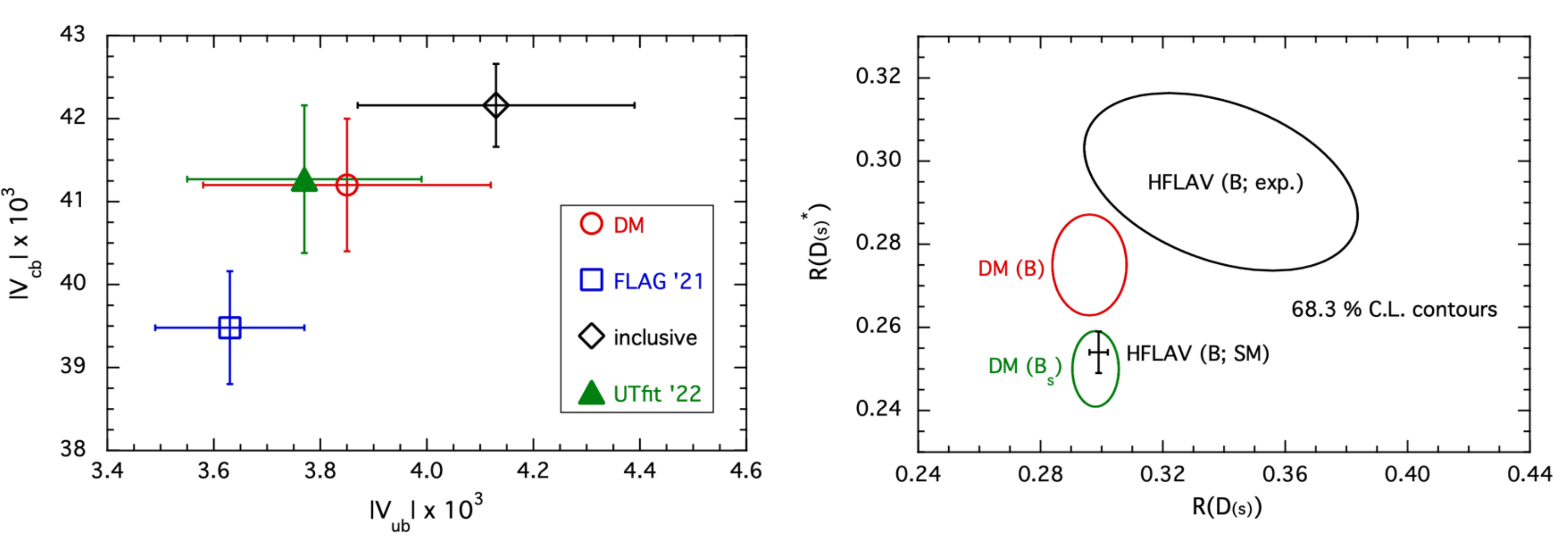}
\vspace{-0.75cm}
\caption{\it \small \textbf{Left panel:} $\vert V_{cb} \vert$ vs $\vert V_{ub} \vert$ plot. The numerical values corresponding to the various entries (DM estimates, FLAG results \cite{FLAG21}, inclusive values \cite{Bordone:2021oof, PDG}, UTfit indirect determinations \cite{UTfit}) can be found in Tab.~\ref{tab1}. \textbf{Right panel:} $R(D_{(s)}^*)$ vs $R(D_{(s)})$ plot. The numerical values corresponding to the various entries (DM estimates, HFLAV experimental average, HFLAV SM average) can be found in Tab.~\ref{tab2}.}
\label{Summary}
\end{figure}

In this contibution we have reviewed the main features of the DM approach, which is an interesting tool to implement unitarity and LQCD calculations in the analysis of exclusive charged-current semileptonic decays of mesons and baryons. In Fig.~\ref{Summary} we have condensed the results obtained so far from the application of the DM method to the semileptonic $B \to D^{(*)}$ \cite{paperoIII, EPJC}, $B_s \to D_s^{(*)}$ \cite{BsDs}, $B \to \pi$ and $B_s \to K$ \cite{paperoV} decays. The DM values of the CKM matrix elements in the left panel represent the averages of all the DM determination of $\vert V_{cb} \vert$ and $\vert V_{ub} \vert$ from the various decay channels. For both the CKM matrix elements, the DM determinations are compatible with the corresponding inclusive values within the $1\sigma$ level. Furthermore, the DM values are practically identical to the indirect determinations coming for the latest analysis by the UTfit Collaboration \cite{UTfit}. The values of the LFU observables (for both the $B \to D^{(*)}$ and the $B_s \to D_s^{(*)}$ decays) are, instead, shown in the right panel, together with the experimental average and the SM one by HFLAV. By using the FNAL/MILC computations of the FFs for the $B \to D^*$ channel, we observe that the tension between theoretical expectations and measurements of $R(D^{(*)})$ is reduced. 

\begin{table}[htb!]
\centering
\caption{Numerical values of the CKM matrix elements $\vert V_{cb} \vert$ and $\vert V_{ub} \vert$ plotted in Fig.~\ref{Summary}.}
\label{tab1}       
{\small
\begin{tabular}{c|c|c|c|l|cc}
& Decay channel & DM values & FLAG '21 & Inclusive & UTfit '22\\\hline
$\vert V_{cb} \vert \times 10^3$ & $B_{(s)} \to D_{(s)}^{(*)}$ & 41.2 (8) & 39.48 (68) & 42.16 (50) & 41.27 (89) \\\hline
$\vert V_{ub} \vert \times 10^3$ & $B_{(s)} \to \pi(K)$  & 3.85 (27) & 3.63 (14) & 4.13 (26) & 3.77 (22) \\
\end{tabular}
}
\end{table}

\begin{table}[htb!]
\centering
\caption{Numerical values of the LFU observables plotted in Fig.~\ref{Summary}.}
\label{tab2}       
{\small
\begin{tabular}{c|c|c|c}
& DM values & HFLAV '21 (exp) & HFLAV '21 (SM)\\\hline
$R(D)$   & 0.296 (8) & 0.339 (26) (14) & 0.299 (3) \\\hline
$R(D^*)$ & 0.275(8) & 0.295 (10) (10) & 0.254 (5) \\\hline
$R(D_s)$ & 0.298 (5) &&\\\hline
$R(D_s^*)$ & 0.250 (6) &&\\
\end{tabular}
}
\end{table}

\end{document}